\documentclass[prd,nofootinbib,twocolumn,superscriptaddress,preprintnumbers,balancelastpage,nofootinbib]{revtex4-1}

\usepackage[dvipdfmx,hiresbb]{graphicx} 
\usepackage{graphicx}
\usepackage{amsmath,amssymb}
\usepackage{ulem} 
\usepackage{epsf}
\usepackage{color}   
\usepackage{here}
\usepackage{endnotes}
\allowdisplaybreaks[1]

\begin{document}
\title{Search for new light particles at ILC main beam dump}  

\author{Yasuhito Sakaki}
\affiliation{
High Energy Accelerator Research Organization (KEK), Tsukuba, Ibaraki 305-0801, Japan
}

\author{Daiki Ueda}
\affiliation{
Department of Physics, Faculty of Science, University of Tokyo, Bunkyo-ku, Tokyo 113-0033, Japan}

\begin{abstract}
We perform a feasibility study of a beam dump experiment at the International Linear Collider (ILC).  To investigate the sensitivity to new light particles at the experiment, we consider models for axion-like particles (ALPs) and a light scalar particle coupled to charged leptons.  For both models, we show that the detection sensitivity is almost an order of magnitude higher than other beam dump experiments in the small coupling region. For ALPs, it is shown that the ILC beam dump experiment is highly complementary to bounds from astrophysics.
In addition, for the model of the scalar particle, the region favored by the muon $g-2$ experiment can be explored.
\end{abstract}

\maketitle

\section{Introduction}
The International Linear Collider (ILC) experiment is one of the next generation experiments using high energy collision with electron ($e^-$) and positron ($e^+$) beams~\cite{ILC}.  It is expected to be used to search for new particles with the electroweak charge and to measure the properties of the Higgs boson precisely.

In the ILC experiment, main beam dumps are expected to be installed for safety.
Almost all $e^+ e^-$ beams that pass the collision point are discarded in the main beam dumps, and photons, electrons, muons, etc. are produced in the electromagnetic shower.
The muons pass through the beam dump due to its strong permeability, and these energies are injected into a muon shield that may be placed behind the beam dump.  If physics beyond the standard model (BSM) predict particles whose interaction is very weak, the injected energy into the beam dump can be converted to the new particles.  It is tempting to plan an experiment to explore these new particles by using the discarded particles after $e^+ e^-$ collisions.

An experiment to detect the signs of new particles with a detector installed behind the beam dump is called a beam dump experiment.
Models containing these particles are attractive candidates for BSM.
For example, the Peccei-Quinn symmetry, which solves the strong CP problem, generates axion\cite{Peccei:1977hh,Peccei:1977ur}.  
Also, light scalar particles that interact with SM lepton\cite{Chen:2015vqy,Batell:2016ove,Krnjaic:2019rsv}
is known as a model that can explain muon $g-2$ experiments\cite{Bennett:2002jb,Bennett:2004pv,Bennett:2006fi}.
Since the new particles contained in these models can have a long lifetime, it is difficult to detect them in the LHC experiment and the ILC main experiment.  Therefore, the beam dump experiment has a complementary role to the collider experiments.

We perform a feasibility study of a beam dump experiment at the ILC, which provides a possibility to search for new light particles.
In this paper, we show the sensitivity for the beam dump experiment using photons, electrons and muons.
As benchmark models, we focus on axion-like particles that interact with photons, and a light scalar particle that interacts with the standard model leptons.
The two main features of the ILC beam dump experiment can be summarized as follows.
The first is that it can be performed in parallel with the main experiment at ILC (the $e^+ e^-$ collision experiment).
Consequently, data can always be acquired in the beam dump experiment while the main experiment is running.
The second is that multiple BSMs can be probed at the same time using the photons, electrons, and muons generated by the electromagnetic shower.
In contrast to Ref.~\cite{Kanemura:2015cxa}, we also take into account the photons, and muons for BSM searches.
Moreover, it is advantageous that the detailed design behind the main beam dump in the ILC experiment is in the planning stage, which allows a high degree of design freedom for the BSM searches.\\

\section{Experimental setup}
\begin{figure*}
\begin{center}
\includegraphics[width=.8\textwidth]{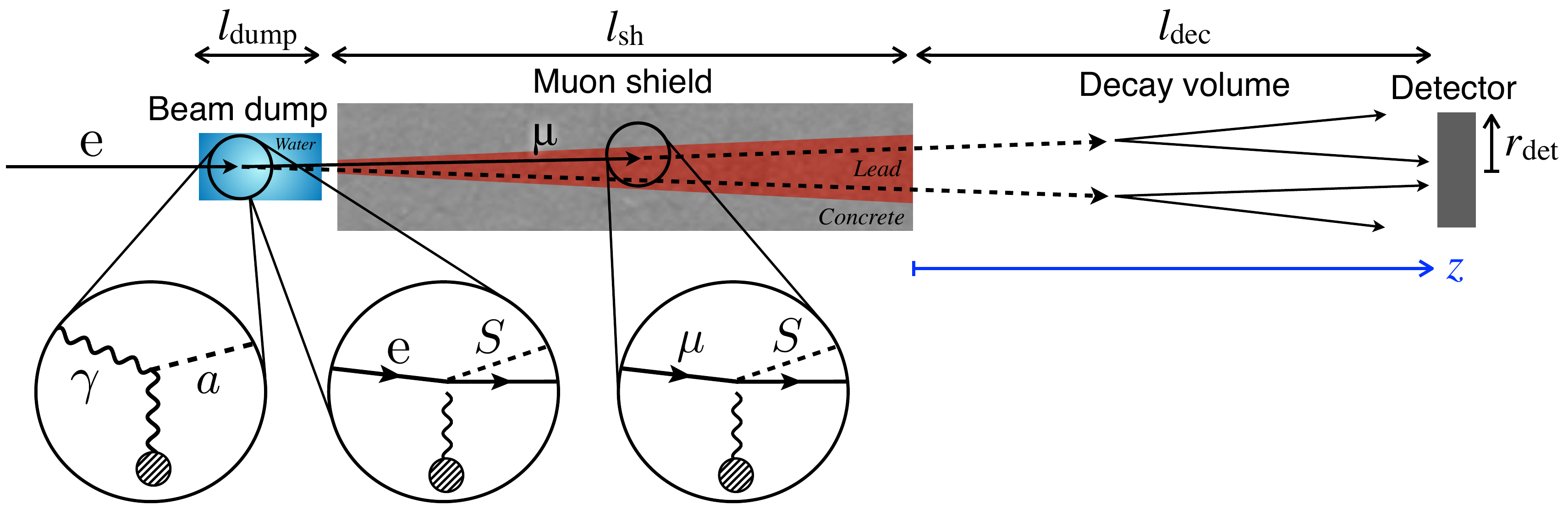}
\caption{An experimental setup consisting of four parts: the main beam dump, a muon shield, a decay volume, and a detector.
}
\label{fig:exp}
\end{center}
\end{figure*}

An experimental setup consists of four parts: the main beam dump, a muon shield, a decay volume, and a detector.
Fig.~\ref{fig:exp} shows the outline of the layout.  Water is planned as the absorber in the main beam dump of ILC~\cite{Satyamurthy:2012zz}.  The length of water cylinder along the beam axis $(l_{\rm dump})$ is approximately 11 m.  In this calculation, the muon shield length $(l_{\rm sh})$ is set to 70 m\footnote{We have confirmed by simulation that this length is sufficient to shield muons and neutrons.} and the decay volume length $(l_{\rm dec})$ is set to 50 m.  It is assumed that lead is placed where muons pass on the shield.  For the convenience of calculation, the shape of the detector is a cylinder, and its axis was aligned with the beam axis.  The radius $(r_{\rm det})$ is set to 2 m and the detection efficiency is assumed to be 100\%. 
We assume that background events can be removed with veto counters located behind the shield and in front of and around the detector.

We consider the case of ILC-250 GeV~\cite{Fujii:2017vwa,Evans:2017rvt} with the beam energy $E_{\rm beam}=125\text{ GeV}$.  The number of incident electrons into the beam dump is assumed to be $N_{\rm EOT}=4\times 10^{21}{\rm /year}$~\cite{ILC}\footnote{In the high luminosity phase of ILC, the number of electrons increases by a factor of about 2, but we do not take this into account in this study.}.   
Inside the main beam dump, the electromagnetic shower produces electrons, positrons, and photons. Then, we consider processes: new scalar particle emissions by the electron interactions with the oxygen nucleus in the beam dump, and axion-like particle emissions by the photons.
In addition, we consider new scalar particle emissions by muon interactions with a lead nuclei in the muon shield.
The muons generated in the forward direction are mainly produced in the muon pair production by a real photon\footnote{The decay of pions coming from the photonuclear reactions also produces muons. We guess the muon pair production by a real photon to have a smaller emission angle and greater angular acceptance.}.
Many muons pass through the beam dump due to their strong penetrating power, and interact with a lead nuclei in the muon shield. \\

\section{Examples of detectable new physics}
%
First, we consider axion-like particles (ALPs) described by the following effective Lagrangian:
\begin{align}
\delta\mathcal{L} =
	-\frac{1}{4}g_{a\gamma\gamma} a F_{\mu\nu}\tilde{F}^{\mu\nu}
	+\frac{1}{2}(\partial_{\mu}a)^2
	-\frac{1}{2}m_a^2a^2,
	\label{eq:L_ALP}
\end{align}
where $a$ is the ALP, $F^{\mu\nu}$ is a strength of the photon field, and $\tilde{F}^{\mu\nu}=\epsilon_{\mu\nu\lambda \rho}F^{\lambda\rho}/2$.
In our evaluation, it is assumed that a coupling $g_{a\gamma\gamma}$ and a ALP mass $m_a$ are independent parameters.
ALP is produced by a photon in the beam dump, as shown in Fig.~\ref{fig:exp}. After passing through the muon shield, ALP decays in the decay volume and emits two photons, which reach the detector and are observed as a signal.

We estimate the number of signals with the following equation:
\small
\begin{align}
N_{\rm signal} &= 
	N_{\rm EOT}
	\int_{m_a}^{E_{\rm beam}} dk
	\int_{0}^{\pi} d\theta_a
	\int_{0}^{l_{\rm dec}} dz\notag
	\\
	&\times
	\frac{N_{\rm avo}X_0}{A} \frac{dl_\gamma}{dk}
	\cdot \frac{d\sigma_{\gamma a}}{d\theta_a}
	\cdot \frac{dP_{\rm dec}}{dz}
	\cdot \Theta(r_{\rm det}-r_{\perp}), \label{eq:nsig_axion}
\end{align}
\normalsize
where
$N_{\rm EOT}(=4\times10^{21}/{\rm year})$ is the number of electrons injected into the beam dump, 
$E_{\rm beam}(=125$ GeV) is the electron beam energy, 
$k$ is the photon energy produced in the electromagnetic shower, 
$z$ is the decay position of the ALP ($z=0$ indicates the beginning of the decay volume, see Fig.~\ref{fig:exp}), 
$N_{\rm avo}$ is the Avogadro constant, 
$X_0$ is the radiation length of water, 
and 
$A$ is an effective mass number of water. 
The photon track length in the beam dump is parameterized as~\cite{Nelson:1968dj}, 
\begin{align}
k \frac{dl_\gamma}{dk} = \frac{0.964u}{-\ln(1-u^2)+0.686u^2-0.5u^4}, 
\end{align}
where $u=k/E_{\rm beam}$.
For the ALP production, we used the following angular differential cross section \cite{Tsai:1986tx,Bjorken:1988as,Dobrich:2015jyk,Dusaev:2020gxi} based on the improved Weizsacker-Williams approximation~\cite{vonWeizsacker:1934nji,Williams:1935dka,Kim:1973he}, 
\begin{align}
\frac{d\sigma_{\gamma a}}{d\theta_a}
	\simeq \frac{\alpha g_{a\gamma\gamma}^2 k^4\theta_a^3}{4t^2} G_2(t),
\end{align}
where $G_2(t) \simeq Z^2
	\left(a^2t/(1+a^2t)\right)^2/
	\left(1+t/d\right)^2$, $Z$ is an effective atomic number of water, 
$a=112 Z^{-1/3}/m_e$, $d=0.164\text{ GeV}^2$, and $m_e$ is the electron mass.  
For the momentum transfer $t$, we use the following approximation to avoid cancellation of significant digits in numerical calculations~\cite{Dusaev:2020gxi}: 
$t\simeq k^2\theta_a^2+m_a^4/(4k^2)$. 
For the form factor $G_2$, we use the combined simple atomic and nuclear form factor, which takes into account the screening effect. 
The decay probability of ALP as a function of $z$ is given by
\begin{align}
\frac{dP_{\rm dec}}{dz} =\frac{1}{l_a}e^{-(l_{\rm dump}+l_{\rm sh}+z)/l_a}. \label{eq:dec}
\end{align}
The decay length is given by $l_a = 64\pi E_a/g_{a\gamma\gamma}^2 m_a^4$.
We use the following approximation for the axion energy~\cite{Dusaev:2020gxi}: $E_a\simeq k-k^2\theta_a^2/(2M_N)-m_a^4/(8M_N k^2)$, 
where $M_N$ is an effective target nucleus mass. 
The function $\Theta$ is the Heaviside step function.

To estimate the angular acceptance, we need the typical deviation of the photon emitted from the axion from the beam axis $(r_{\perp})$.
We estimate the deviation as
\begin{alignat}{4}
r_{\perp} &\sim f_{\rm unc} \sqrt{\sum_{i=1}^3 r_{\perp,i}^2}
								  , \quad &r_{\perp,i}&=\theta_i l_i, \label{eq:rto}
\end{alignat}
where $\theta_1 =8\times10^{-3}{\rm GeV}/k$, $\theta_2 =\theta_a $, $\theta_3 =m_a/k $, $l_1=l_2=l_{\rm dump} + l_{\rm sh} + l_{\rm dec}$, and $l_3 = l_{\rm dec}-z$.
$\theta_1$ is the expected angle of photon in the electromagnetic shower\footnote{
This is estimated in a Monte Carlo simulation using {\tt EGS5}~\cite{Hirayama:2005zm} implemented in {\tt PHITS}~\cite{Sato:2018}.
}, 
$\theta_2$ is the ALP production angle, 
$\theta_3$ is the expected decay angle of photon from the ALP.\footnote{
Before the electrons and positrons are injected into the beam dump, these are swept with magnets to reduce the heat load on the beam dump window, then have an angle. The size of that angle is $\theta\sim 6\times 10^{-4}$~\cite{Yu}, and this effect is negligible in the current experimental setup.
}
Moreover, $l_i$ represents the distance from the point where the production or decay labeled 1-3 above occurred to the place where the detector is located. 
In our evaluation, we calculated the final deviation by summing the squares of these deviations as in the above equation.  
The uncertainty to this estimate is represented by $f_{\rm unc}$, which is expected to have a value on the order of $\mathcal{O}(1)$. 
We set $f_{\rm unc}=1.5$ in this calculation. 
In this setting, our calculation perfectly reproduces a result for an electron beam dump experiment (E137 experiment~\cite{Bjorken:1988as}) in a previous study~\cite{Dolan:2017osp}.  
We will use results summarized in \cite{Dolan:2017osp} as a criterion for the other experimental results in Fig.~\ref{fig:contour_ALP} in the next section, so taking $f_{\rm unc}$ in this way allows for a fair comparison of our ILC calculations with the results of the other experiments.


Next, we consider the following Lagrangian for a scalar particle coupled to the charged leptons in the standard model (SM):
\begin{align}
\delta\mathcal{L} =
	\frac{1}{2}(\partial_{\mu}S)^2
	-\frac{1}{2}m_S^2S^2
	-\sum_{\ell=e,\mu,\tau} g_{\ell} S \bar{\ell}\ell,
	\label{eq:L_S}
\end{align}
where $S$ is a new scalar particle, and $g_\ell$ is a coupling between $S$ and the SM charged leptons.
Then we assume two models as a benchmark:
\begin{alignat}{2}
& g_{\ell}\propto m_{\ell}, \quad &\text{(Model A)} \label{eq:modelA}\\
& g_{\mu}\neq 0,~ g_e=g_{\tau}=0, \quad &\text{(Model B)} \label{eq:modelB}
\end{alignat}

As shown in Fig.~\ref{fig:exp}, the scalar particle can be produced in bremsstrahlung from electron and muon. 
The generated scalar particle decays into photons\footnote{In order to clarify the sensitivity of the muon to the light scalar particle, we assume that the interaction between $S$ and photon is not generated at the tree level. Then, $S$ decays into photons because of the muon loop diagram.}, electron-positron, and muon pair in the decay volume, which reach the detector and become a signal. 
The decay channel depends on the model and the mass of the scalar particle. 
The method of evaluating the number of these signals is summarized in Appendix~\ref{app:NS}.\\

\section{Results}
%
\begin{figure}[!t]
\begin{center}
\includegraphics[width=7.0cm, bb=0 0 550 520]{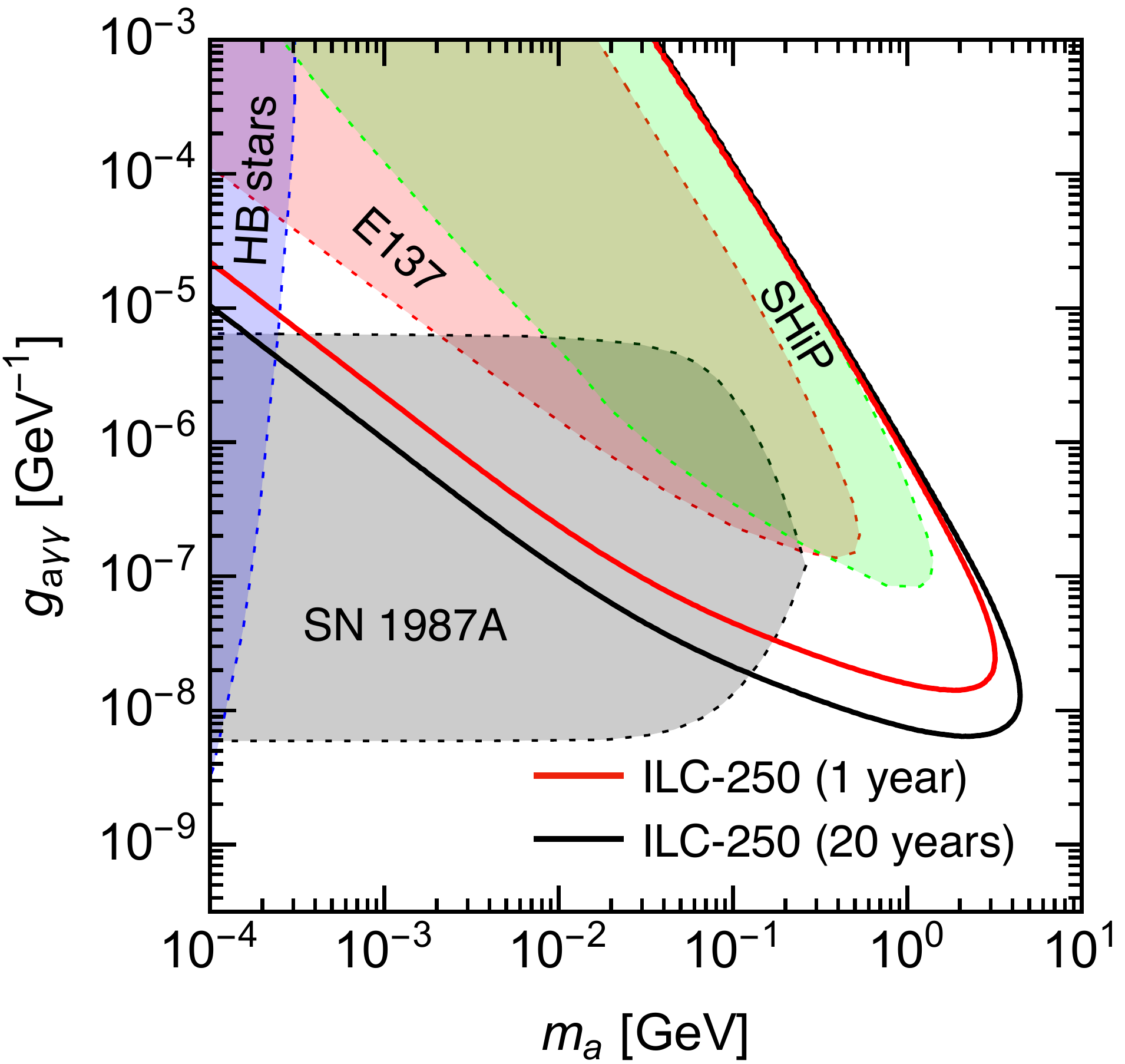}
\caption{{
The red and black curves show the bounds of sensitivity for ILC-250 GeV at 95\% C.L. with 1- and 20-year statistics.
The shaded regions are constraints for E137 from \cite{Dolan:2017osp}, SN 1987A from \cite{Dolan:2017osp,Jaeckel:2017tud}, HB stars from \cite{Cadamuro:2011fd}, and SHiP from \cite{Dolan:2017osp,Dobrich:2015jyk,Dobrich:2019dxc}.  
}}
\label{fig:contour_ALP}
\end{center}
\end{figure}
In Fig.~\ref{fig:contour_ALP}, 
the red and black curves show the bounds of sensitivity for ILC-250 GeV at 95\% C.L. with 1- and 20-year statistics.
The shaded regions are constraints from E137~\cite{Bjorken:1988as}, SN 1987A~\cite{Jaeckel:2017tud}, HB stars~\cite{Cadamuro:2011fd}, and SHiP~\cite{Anelli:2015pba}. 
It can be seen that the ILC beam dump experiment has almost an order of magnitude higher sensitivity than other beam dump experiments in the small coupling region. 
In addition, the ILC beam dump experiment is highly complementary to bounds from astrophysics.

For ease of understanding Eq.~(\ref{eq:nsig_axion}), we discuss the dependence of the parameters on the curves using some approximations.
Let us consider the case of $l_a \gg l_{\rm sh}$, where the ALP has a longer lifetime, and the decay probability in Eq.~(\ref{eq:dec}) is roughly equal to $l_a^{-1}$. 
The parameter dependence of the curves is divided into two cases. Case I: $m_a<\mathcal{O}(10^{-2})\,{\rm GeV}$.
Then, $\theta_1$ tends to be greater than $\theta_2$ and $\theta_3$, and the minimum value of photon energy is approximately $k_{\rm min}\sim 8\times 10^{-3}\,{\rm GeV}\times (l_{\rm dump}+l_{\rm sh}+l_{\rm dec})/r_{\rm det}$ according to Eq.~\eqref{eq:rto}.
Case II: $m_a>\mathcal{O}(10^{-2})\,{\rm GeV}$, where $\theta_3$ tends to be greater than $\theta_1$ and $\theta_2$.
Then, the minimum value of photon energy is approximately proportional to $m_a$ ($k_{\rm min}\propto m_a$) according to Eq.~\eqref{eq:rto}.
Consequently, the number of signals is approximately integrated as
\small
\begin{align}
N_{\rm signal} &\sim 
	\left( \frac{E_{\rm beam}}{\text{125 GeV}} \right)
	\left( \frac{N_{\rm EOT}}{4\times10^{21}} \right)
	\left( \frac{g_{a\gamma\gamma}}{2\times10^{-7}\,{\rm GeV}^{-1}} \right)^4\notag
	\\
	& \times
	\left( \frac{r_{\rm det}}{2\,{\rm m}} \right)^2 
	\left( \frac{l_{\rm dec}}{50\,{\rm m}} \right)
	\left( \frac{ l_{\rm dump}+l_{\rm sh}+l_{\rm dec}}{ 121\,{\rm m}} \right)^{-2}\notag
	\\
	& \times
	\begin{cases}
	\left( \frac{m_a}{10^{-2}\,{\rm GeV}} \right)^4,~({\rm Case~I})\\
	\left( \frac{m_a}{10^{-2}\,{\rm GeV}} \right)^2,~({\rm Case~II})
	\end{cases}
	\label{eq:Nsig_App}
\end{align}
\normalsize
%
%
where we use following approximations: $dl_{\gamma}/dk\propto E_{\rm beam}/k^2$, 
$E_a \simeq k$, and $dP_{\rm dec}/dz \simeq 1/l_a$, and neglect the logarithmic dependence of $k$ in $\sigma_{\gamma a}$.
Eq.~\eqref{eq:Nsig_App} shows the parameter dependence on the lower side of the contour in Fig.~\ref{fig:contour_ALP}.  
The sensitivity to small coupling can be maximized by setting to
\begin{align}
l_{\rm dec}=l_{\rm dump}+l_{\rm sh}.
\end{align}
By using Eq.~(\ref{eq:Nsig_App}) and E137 setup $(E_{\rm beam}=20~{\rm GeV}, N_{\rm EOT}\simeq 2\times 10^{20}, l_{\rm dump}+l_{\rm sh}\simeq 179~{\rm m}, l_{\rm dec}=204~{\rm m}, r_{\rm det}\simeq 1.5~{\rm m})$, it becomes clear that ILC is more sensitive to coupling 5-10 times smaller than E137 for a given $m_a$.

Next, consider the case of $l_a \ll l_{\rm sh}$, where most ALPs decay in the shield. The sensitivity is determined by the exponential factor in Eq.~\eqref{eq:dec}. 
The upper side of the contour in Fig.~\ref{fig:contour_ALP} corresponds to this case, which is characterized by the following equation: 
\begin{align}
g_{a\gamma\gamma}^2 m_a^4
	\frac{l_{\rm dump}+l_{\rm sh}}{E_{\rm beam}}
	\sim {\rm Const.}
\end{align}
By shortening the length of the beam dump and the muon shield $(l_{\rm dump}+l_{\rm sh})$, the probability that ALP decays in front of the decay volume can be reduced, and the sensitivity region enlarge to the upper right. (See in Appendix~\ref{app:lsh}.)
Moreover, the sensitivity can be enlarged using higher energy beam $(E_{\rm beam})$ because the higher energy ALP has a longer lifetime by a larger boost factor.

Fig.~\ref{fig:contour_ALP} demonstrates that ILC has better sensitivity than SHiP.
In a proton beam dump experiment, the ALP is generated in the Primakov production as in an electron beam dump experiment. The main source of the photons in the initial state would come from meson decays.
We guess that photons generated by meson decays have much larger angles compared to the bremsstrahlung photon from the electron, therefore the number of photons reaching the detector is reduced in the proton beam dump experiment.


\begin{figure}[!t]
\begin{center}
\includegraphics[width=7.0cm, bb=0 0 550 560]{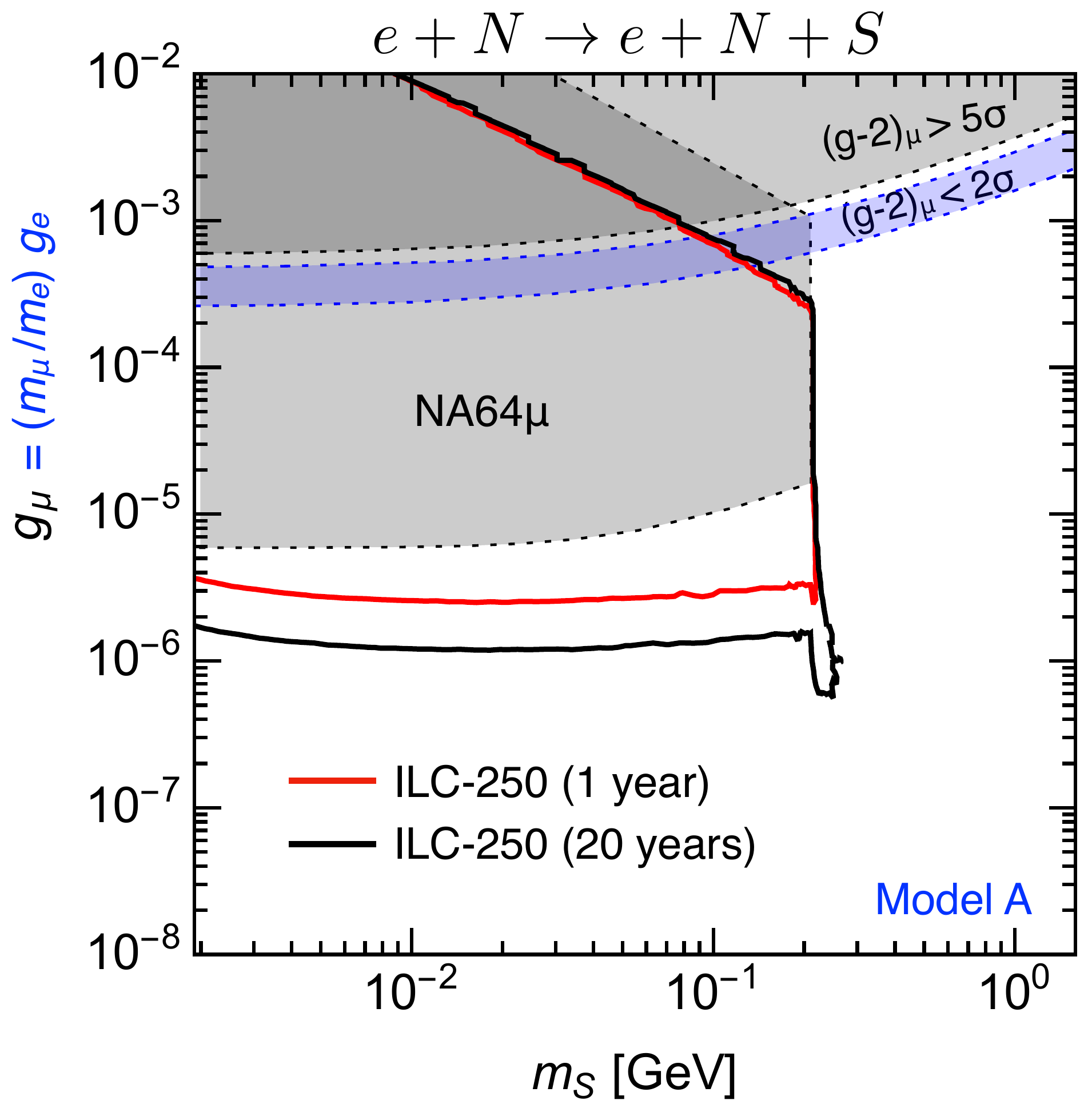}
\includegraphics[width=7.0cm, bb=0 0 550 560]{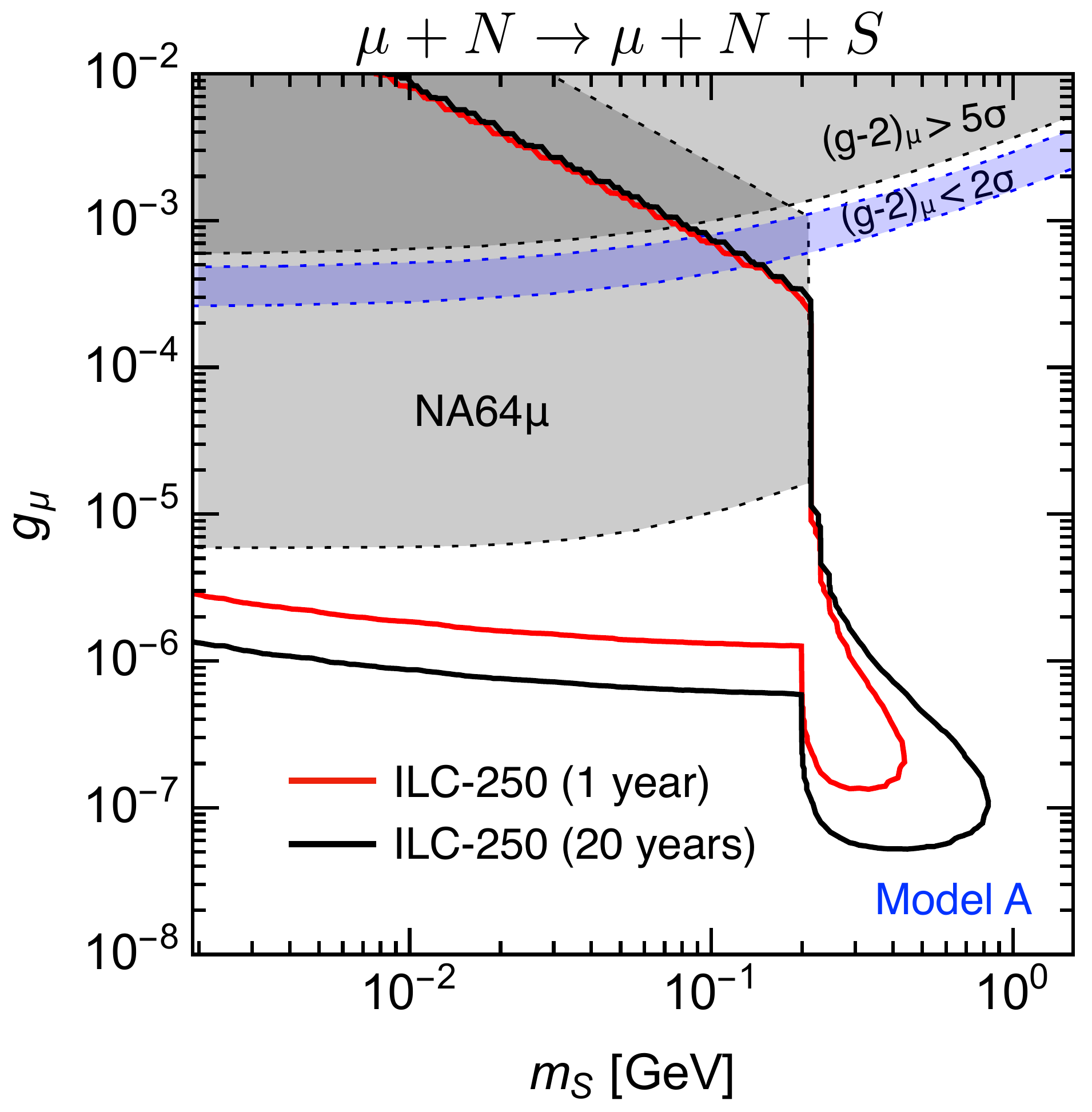}
\caption{{ 
The top (bottom) figure is the result of a process containing an electron (muon) in the initial state.
The gray shaded regions are a constraint from NA64$\mu$ and muon $g-2$ from \cite{Chen:2017awl}. Note that, although the results in $m_S>2m_{\mu}$ are absence for NA64$\mu$, it would also have a sensitivity in that region generally.
}}
\label{fig:contour_SA}
\end{center}
\end{figure}

\begin{figure}[!t]
\begin{center}
\includegraphics[width=7.0cm, bb=0 0 550 560]{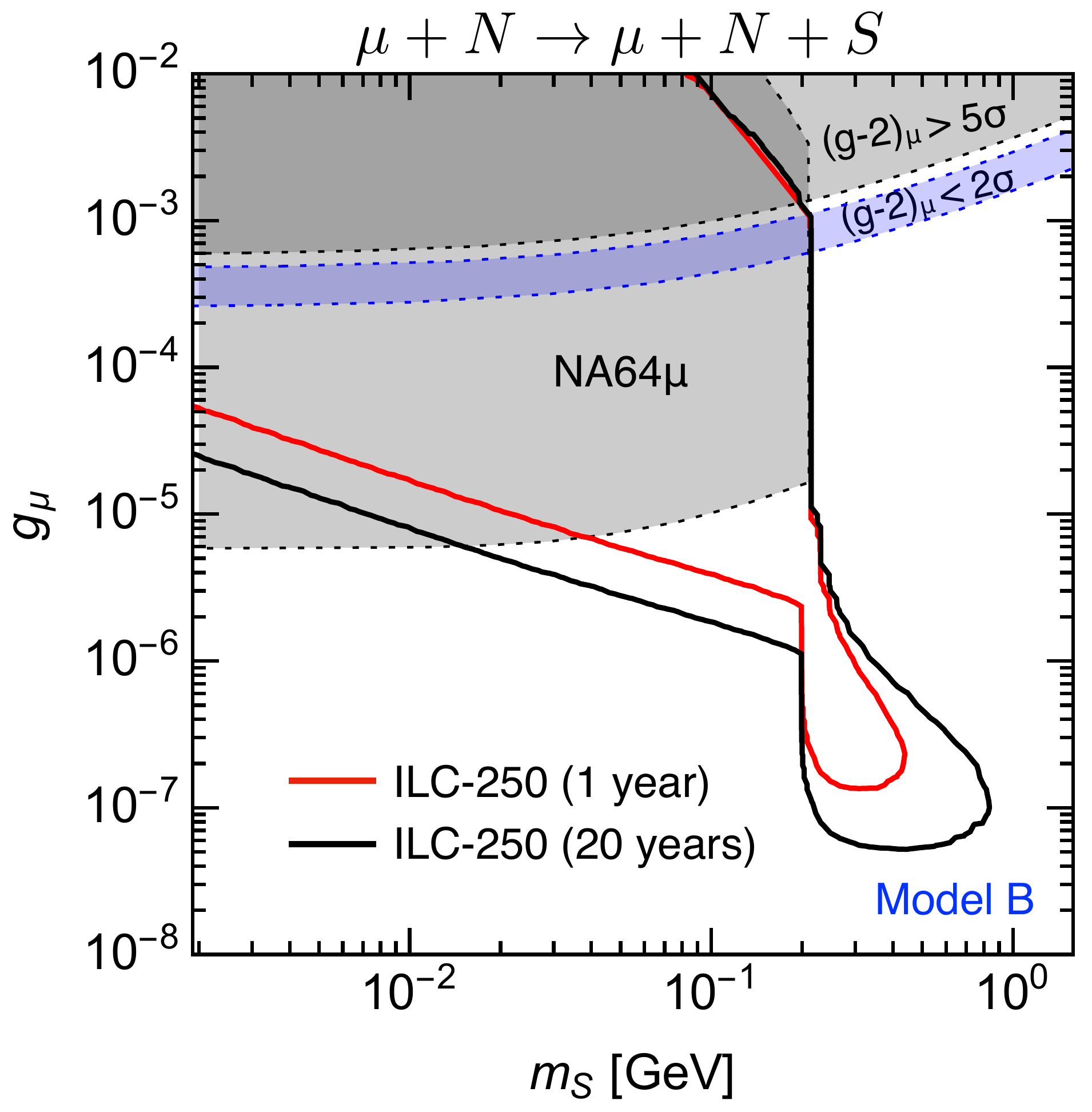}
\caption{{ 
The same plot as Fig.~\ref{fig:contour_SA} but for the Model B.
}}
\label{fig:contour_SB_mu}
\end{center}
\end{figure}

Now, let us move on sensitivities for the light scalar particle introduced in Eq.~(\ref{eq:L_S}).
In Fig.~\ref{fig:contour_SA}, 
the red and black curves show the bounds of sensitivity for ILC-250 GeV at 95\% C.L. with 1- and 20-year statistics.
The results for Model A are shown.  
The top (bottom) figure is the result of a process containing an electron (muon) in the initial state.
The gray shaded regions are the constraint from muon $g-2$ and the sensitivity from NA64$\mu$~\cite{Andreas:2013lya,Gninenko:2014pea,Chen:2017awl}.
The blue band region is the favored region for muon $g-2$ at 2 sigma level.
From Fig.~\ref{fig:contour_SA}, it can be seen that the ILC beam dump experiment has high sensitivity in the small coupling region.  
In addition, the ILC beam dump experiment has sensitivity to the region favored by muon $g-2$, similar to the NA64$\mu$ experiment, which is an experiment to measure missing energy using an {\it advanced} technology.  
Thus, if the experiment detects a sign of new physics in this region, it can be verified by the configuration of this ILC beam dump experiment using the {\it developed} technology for visible particle searches.

Fig.~\ref{fig:contour_SB_mu} is the same plot as Fig.~\ref{fig:contour_SA} but for Model B.  
In this model, only results are shown for the case where the muon is the initial state, since the scalar particle has only the Yukawa coupling with muons.
In this case, the ILC beam dump experiment can cover most of the region favored by muon $g-2$ in $m_S<2m_{\mu}$.

Here, we discuss the difference between the ILC beam dump experiment and the missing energy search such as NA64$\mu$ and ${\rm M}^3$ experiment~\cite{Kahn:2018cqs}. In $m_S<2m_{\mu}$, the ILC beam dump experiment detect the photons as signal events, and the interaction between $S$ and photon can control the sensitivity of the experiment. In order to clarify the sensitivity of the initial muon to the light particle search, we assume that the interaction between $S$ and photon is only generated by the muon loop diagram. Then, the decay width of $S$ is summarized in \cite{Chen:2017awl} as 
\begin{align}
\Gamma_{\gamma\gamma}=\frac{\alpha^2 m_S^3}{64\pi^3} \left|\sum_{l=e,\mu,\tau}\frac{g_l}{m_l}\tau_l \left[1+(1-\tau_l)f(\tau_l) \right] \right|^2,
\end{align}
with $\tau_l =4m_l^2/m_S^2$, and
\begin{align}
f(\tau)=	\begin{cases}
	\arcsin^2 (\tau^{-1/2}),~~~\tau >1\\
	-\frac{1}{4}\left[\ln \left(\frac{1+\sqrt{1-\tau}}{1-\sqrt{1-\tau}}\right)-i\pi\right]^2,~~~\tau \leq 1.
	\end{cases}
\end{align} 
On the other hand, in NA64$\mu$ and ${\rm M}^3$ experiment, the sensitivity is independent of the interaction between $S$ and photon because of the missing energy search.

Here, we comment about the parameter dependence on the lower side of contours.
Similar to the ALPs, the cases correspond to $l_{\rm sh} \ll l_S$.
For Model A corresponding to Fig.~\ref{fig:contour_SA}, the number of signals is proportional to $\sigma_{S}\cdot l_S^{-1}\propto g_{\mu}^4 m_S^2/E_{S,{\rm min}}^2$ by using $\sigma_{S}\propto g_{\mu}^2$ and $l_S^{-1}\propto g_{\mu}^2 m_S^2/E_{S}$, where $\sigma_S$ is the production cross section for the scalar particle, $l_S$ is the decay length of the scalar particle, and $E_{S,{\rm min}}$ is the minimum energy of the scalar particle that contributes to the number of signals. 
$E_{S,{\rm min}}$ is approximately proportional to $m_S$ according to Eq.~\eqref{eq:thap}.
Consequently, the number of signals is proportional to $g_{\mu}^4 $, where the $m_S$ dependence vanishes.
For Model B corresponding to Fig.~\ref{fig:contour_SB_mu}, $l_S^{-1}\propto g_{\mu}^2 m_S^4/E_{S}$ because of decays into photons, and the number of signals is proportional to $g_{\mu}^4  m_S^2$.

Last, we comment on a structure of the contours around $m_S=2m_{\mu}$.  
When $m_S$ becomes larger than $2m_{\mu}$, the decay mode into a muon pair opens, and its decay length suddenly shortens.
Therefore, the decay particles from the scalar particle tend to be stopped in the muon shield.
However, if the coupling becomes smaller, the probability of decaying particles passing through the muon shield increases. So a constraint region appears in the smaller coupling region and in $m_S>2m_{\mu}$.\\

\section{Summary}
\label{sec:Summary}
We performed a feasibility study of an experiment using the main beam dump at ILC.
This experiment can be performed in parallel with the main experiment at the ILC using $e^+e^-$ collisions, and almost all $e^+e^-$ beams can be used for the beam dump experiment.
An experimental setup of approximately 130 m in length was considered, including a main beam dump, muon shield, decay region, and detector.
Electrons, photons, and muons generated by the electromagnetic shower in the beam dump were used in the initial state of new physics processes.
To investigate the sensitivity to new light particles at the experiment, we considered models for axion-like particles (ALPs) and a light scalar particle coupled to charged leptons.
We considered the Primakov process for the ALPs and bremsstrahlung from electron or muon for the scalar particle.
We have shown that the sensitivity to both models is almost an order of magnitude higher than other beam dump experiments in the small coupling region.
For the ALPs, it was shown that the ILC beam dump experiment has sensitivity an unexplored region between the other beam dump experiment and bounds from astrophysics.
In addition, for the model of the scalar particle, it was shown that the region favored by the muon $g-2$ experiment can be explored.

\subsection*{Acknowledgement}
We would like to thank Toshiya Sanami and Yoshihito Namito for giving us an opportunity to work with the ILC beam dump. YS also wish to thank Nobuhiro Terunuma, Yu Morikawa, Yuji Kishimoto and Hirohito Yamazaki for helpful discussions and comments.


\appendix
\section{Number of signals from light scalar particle}
\label{app:NS}

%
%
\label{app:S_eq}
We describe the formula to evaluate the number of signals induced by the light scalar particle.
As shown in Fig.~\ref{fig:exp}, the scalar particle can be produced in bremsstrahlung from electron and muon. The generated scalar particle decays into photons, electron-positron, and muon pair in the decay volume, which reach the detector and become a signal. The decay channel depends on the model and the mass of the scalar particle.

For the scalar particle bremsstrahlung from electron, the number of signals is estimated by
\small
\begin{align}
&N_{{\rm signal},e} =
	N_{\rm EOT}
	\int_{m_{e}}^{E_{\rm beam}} dE_{e}
	\int_{m_S}^{E_{e}-m_{e}} dE_S
	\int_{0}^{\pi} d\theta_S
	\int_{0}^{l_{\rm dec}} dz \nonumber\\
	&\hspace{20pt}\times
	         n_{\rm atom}\frac{dl_{e}}{dE_{e}}
	\cdot \frac{d\sigma_{e,S}}{dE_S \, d\theta_S}
	\cdot \frac{dP_{{\rm dec},e}}{dz}
	\cdot \Theta(r_{\rm det}-r_{\perp,e}).
\end{align}
\normalsize
For muon, it is estimated by
\small
\begin{align}
&N_{{\rm signal},\mu} =
	N_{\rm EOT}
	\int_{m_{\mu}}^{E_{\rm beam}} dE_{\mu_0}
	\int_{m_{\mu}}^{E_{\mu_0}} dE_{\mu}
	\int_{-\pi}^{\pi} d\theta_{\rm MCS} \nonumber\\
	&\hspace{20pt}\times
	\int_{m_S}^{E_{\mu}-m_{\mu}} dE_S
	\int_{0}^{\pi} d\theta_S
	\int_{0}^{l_{\rm dec}} dz
	\frac{dY_{\mu_0}}{dE_{\mu_0}}
	\cdot n_{\rm atom}\frac{dl_{\mu}}{dE_{\mu}} \nonumber\\
	&\hspace{20pt}\times
	\cdot \frac{dP_{\rm MCS}}{d\theta_{\rm MCS}}
	\cdot \frac{d\sigma_{\mu,S}}{dE_S \, d\theta_S}
	\cdot \frac{dP_{{\rm dec},\mu}}{dz}
	\cdot \Theta(r_{\rm det}-r_{\perp,\mu}).
\end{align}
\normalsize
The energy distribution of the muon yield per incident electron behind the beam dump is~\cite{Sakaki:2020cux}
\small
\begin{align}
\frac{dY_{\mu_0}}{dE_{\mu_0}} = 
	\frac{0.572 E_{\rm beam}}{\ln(183Z^{-1/3})}
	\left(\frac{m_e}{m_{\mu}}\right)^2
	\left(\frac{1}{E_{\mu_0}^2}-\frac{1}{E_{\rm beam}^2}\right).
\end{align}
\normalsize
The atomic density of target is $n_{\rm atom}=N_{\rm avo}\rho/A$, 
$\rho$ and $A$ are the mass density and the (effective) mass number of target, where 
the target is water (lead) for the scalar particle bremsstrahlung from electron (muon).
The track length is given by~\cite{Tsai:1986tx}
\small
\begin{align}
\frac{dl_{e}}{dE_{e}} &= 
	\frac{X_0}{\rho}
	\frac{1}{E_{\rm beam}}
	\int_0^{\frac{l_{\rm dump}\rho}{X_0}}dt 
	\frac{[\ln(E_{\rm beam}/E_e)]^{4t/3-1}}{\Gamma(4t/3)}, \\
\frac{dl_{\mu}}{dE_{\mu}} &= \langle dE/dx \rangle_{\rm Lead}^{-1},
\end{align}
\normalsize
where $X_0$ is the radiation length of water.  
Since energy dependence of stopping power is small at higher energy than minimum ionizing energy, we use the following stopping power for lead that is energy independent: $\langle dE/dx \rangle_{\rm Lead}=$ 0.02 GeV/cm. 
The angular distribution stemming from the multiple Coulomb scattering (MCS) of the muon is given by~\cite{Lynch:1990sq,Tanabashi:2018oca}, 
\small
\begin{align}
&\frac{dP_{\rm MCS}}{d\theta_{\rm MCS}} =
	\frac{1}{\sqrt{2\pi\theta_0^2}} \exp\left(-\frac{\theta_{\rm MCS}^2}{2\theta_0^2}\right),\\
&\theta_0 =
	\frac{13.6~{\rm MeV}}{\beta p} 
	\sqrt{\frac{\delta_{\mu}}{x_0^{\rm (Lead)}\beta^2}}
	\left[1+0.038\ln\left(\frac{\delta_{\mu}}{x_0^{\rm (Lead)}\beta^2}\right)\right],\\
&\beta = p/E_{\mu_0}, \quad p=\sqrt{E_{\mu_0}^2-m_{\mu}^2}, \quad x_0^{\rm (Lead)}=0.56\,{\rm cm}.
\end{align}
\normalsize
The distance that muon passes through the muon shield before emitting the scalar particle is
\small
\begin{align}
\delta_{\mu} = \frac{E_{\mu_0}-E_{\mu}}{\langle dE/dx \rangle_{\rm Lead}}.
\end{align}
\normalsize
The energy-angle production cross section for the scalar particle is given with Weizsacker-Williams approximation as~\cite{vonWeizsacker:1934nji,Williams:1935dka,Liu:2016mqv}
\small
\begin{align}
&\frac{d\sigma_{i,S}}{dE_S \, d\theta_S} =
	\frac{g_i^2\alpha^2}{4\pi}
	\frac{\sin\theta_S}{E_i-E_X}
	\sqrt{\frac{E_S^2-m_S^2}{E_i^2-m_i^2}}
	\frac{\mathcal{A}}{2\tilde{t}_{\rm min}}
	\chi,\\
&\mathcal{A} =
	\frac{x^2}{1-x}+2(m_S^2-4m_i^2)\frac{\tilde{u}x+m_S^2(1-x)+m_i^2 x^2}{\tilde{u}^2},\\
&\tilde{t}_{\rm min}=\frac{\tilde{u}^2}{4(1-x)^2 E_i^2}, \quad
	\tilde{u}=-xE_i^2\theta_S^2-m_S^2\frac{1-x}{x}-m_i^2x,\\
&x=E_S/E_i, \quad
	\chi \simeq \int_{m_S^4/(4E_i^2)}^{m_S^2+m_i^2} dt \frac{t-m_S^4/(4E_i^2)}{t^2} G_2(t),
\end{align}
\normalsize
where $Z$ in $\chi$ is an (effective) atomic number of each target. 
The decay probability of the scalar particle is given by
\small
\begin{align}
&\frac{dP_{{\rm dec},e}}{dz}     = \frac{1}{l_S}e^{-\frac{l_{\rm dump}+l_{\rm sh}+z}{l_S}},\\
&\frac{dP_{\rm dec,\mu}}{dz} = \frac{1}{l_S}e^{-\frac{l_{\rm sh}-\delta_{\mu}+z}{l_S}},
\end{align}
\normalsize
where the decay length of the scalar particle $(l_S)$ is summarized in Sec.~2 of \cite{Chen:2017awl}.

To estimate the angular acceptance, we need to know a typical deviation from the beam axis $(r_{\perp})$ for the visible particles emitted from the scalar particle.
We estimate it by
\small
\begin{align}
r_{\perp,i} &\sim f_{\rm unc} \sqrt{\sum_{j=1}^3 r_{\perp,i,j}^2}, \quad
	 r_{\perp,i,j}=\theta_{i,j} l_{i,j}, \quad i=e,\mu,\label{eq:thap}\
\end{align}
\normalsize
where
\small
\begin{align}
\theta_{e,1} &=16\times10^{-3}{\rm GeV}/E_e,\\
\theta_{\mu,1} &=\sqrt{\left(\frac{2m_{\mu}}{E_{\mu_0}}\right)^2+\theta_{\rm MCS}^2},\\
\theta_{i,2} &=\theta_S,\\
\theta_{i,3} &=\frac{m_S}{E_S} \left( 1-\frac{4m_F^2}{m_S^2} \right)^{1/2} \left( 1-\frac{4m_F^2}{E_S^2} \right)^{-1/2},\\
l_{e,1} &= l_{e,2} = l_{\rm dump} + l_{\rm sh} + l_{\rm dec},\\
l_{\mu,1} &= l_{\rm sh} + l_{\rm dec},\\
l_{\mu,2} &= l_{\mu,1}-\delta_{\mu},\\
l_{i,3} &= l_{\rm dec}-z.
\end{align}
\normalsize
$\theta_{e,1}$ is the expected angle of electron in the electromagnetic shower estimated in MC simulation,
$\theta_{\mu,1}$ is a combined angle between the muon production angle and the deviation from the multiple Coulomb scattering,
$\theta_{i,2}$ is the production angle of the scalar particle, 
$\theta_{i,3}$ is a typical angle of decay particles from the scalar particle,
$m_F$ is the mass of the decay particles from the scalar, 
$l_i$ is the distance from the point where the production or decay labeled 1-3 above occurred to the place where the detector is located. 
The function $\Theta$ is the Heaviside step function.

\section{Shield length dependence of sensitivity for light scalar particle}
\label{app:lsh}
\begin{figure}[!t]
\begin{center}
\includegraphics[width=6.0cm, bb=0 0 550 560]{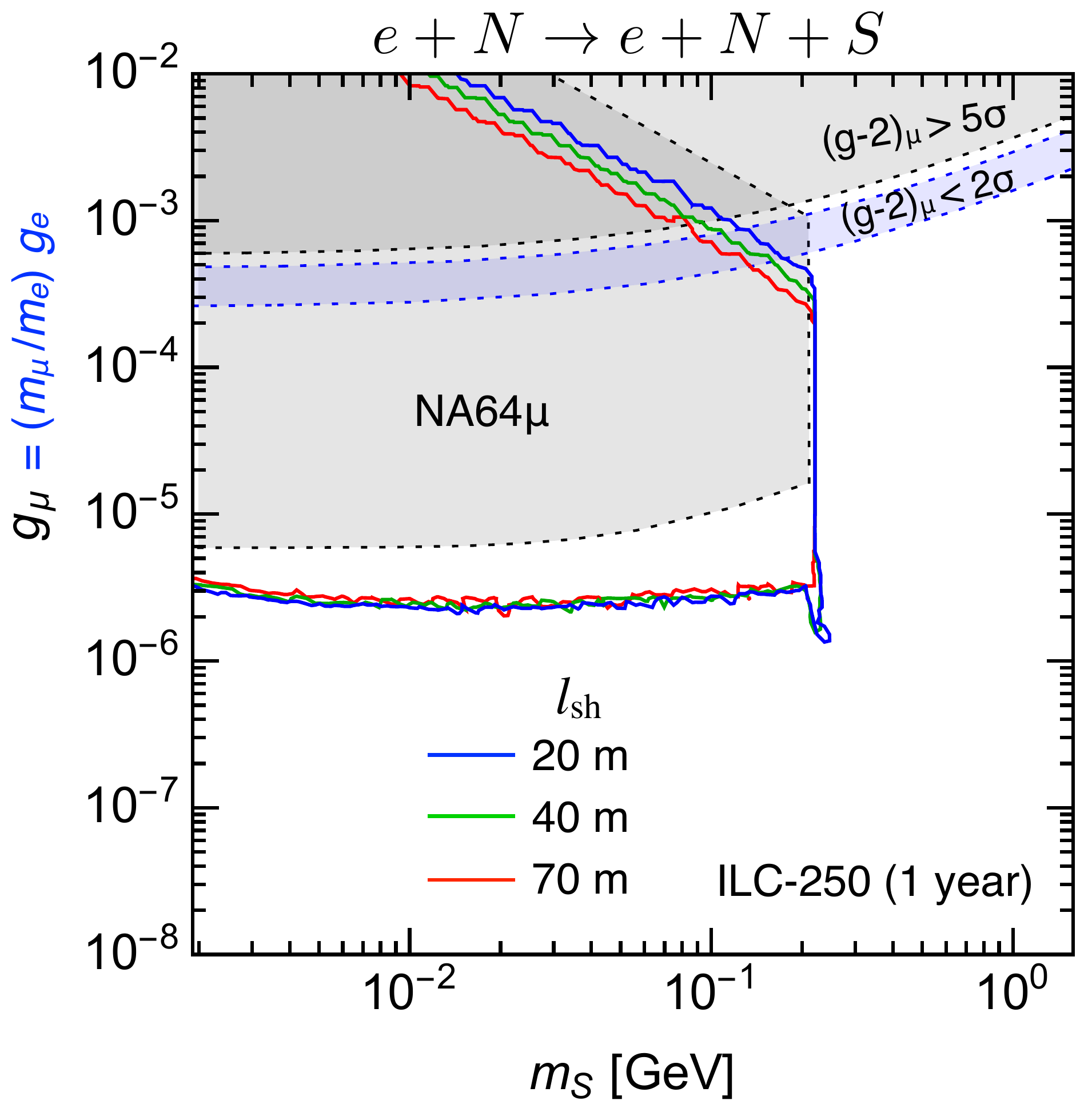}
\quad
\includegraphics[width=6.0cm, bb=0 0 550 560]{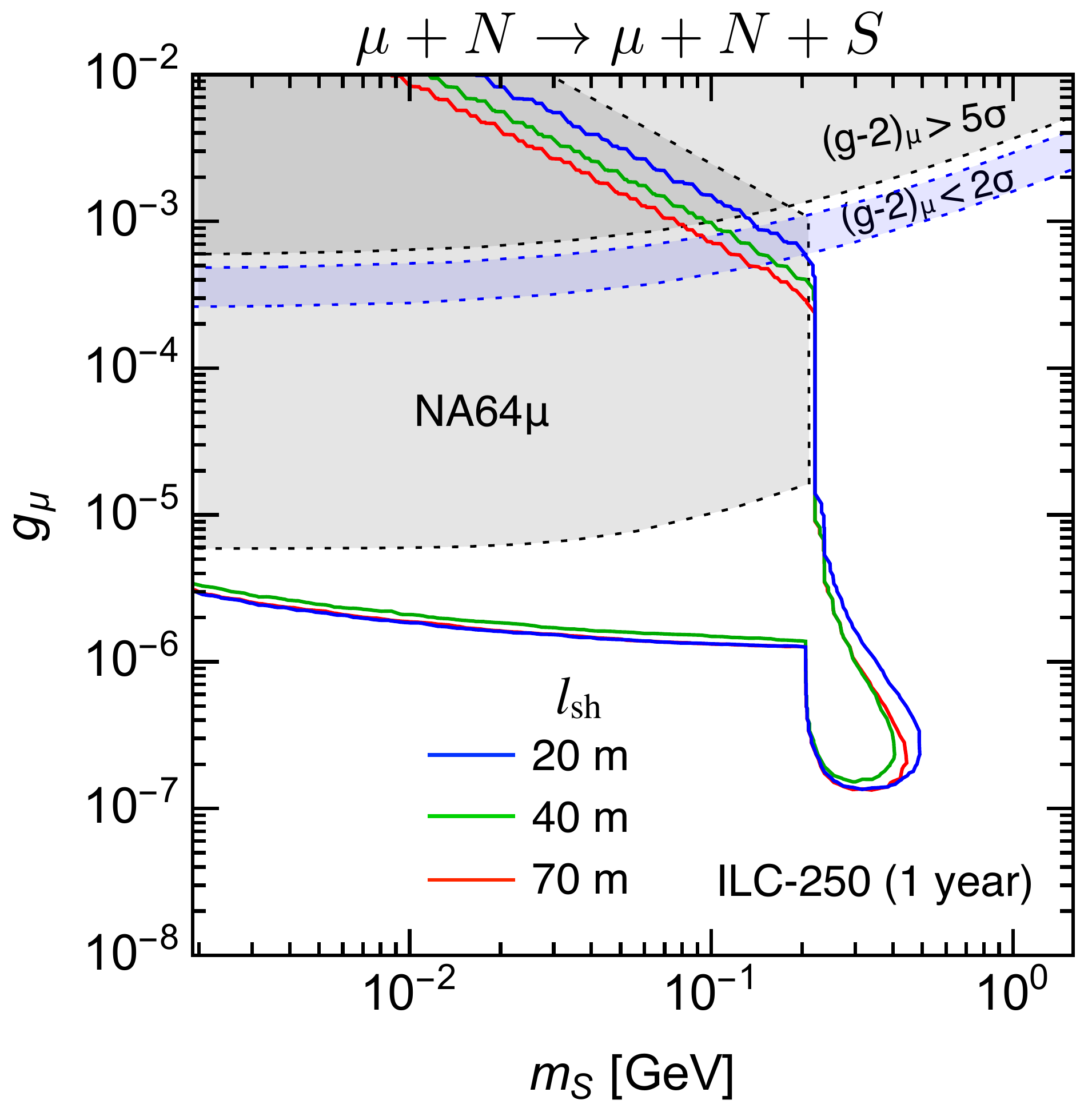}
\caption{{\footnotesize 
The same plot as Fig.~\ref{fig:contour_SA} but results for different shield lengths, 20, 40, and 70 m are compared.
}}
\label{fig:contour_SA_lsh}
\end{center}
\end{figure}

In Fig.~\ref{fig:contour_SA_lsh}, the results are shown in the case that the shield length is 20,  40, and 70 m. By shortening the muon shield, it is possible to enlarge the constraint to the region favored by muon $g-2$.  
This is because the probability that light scalar particle decays in front of the decay volume can be reduced, and the sensitivity region enlarge to the upper right.


\end{document}